\documentclass[achemso,twocolumn,superscriptaddress,floatfix,amsmath,amssymb]{revtex4-2}
\usepackage{amsmath}
\usepackage{amssymb}
\usepackage{braket}
\usepackage{graphicx}
\usepackage[table]{xcolor}
\usepackage{verbatim}
\usepackage{float}
\usepackage{color}
\usepackage{siunitx}
\usepackage{gensymb}
\usepackage{bm}
\usepackage{xcolor}
\usepackage{multirow}
\usepackage{footnote}
\usepackage{bbm}
\usepackage{enumitem}
\usepackage{oplotsymbl}
\usepackage{bbding}
\usepackage{hyperref}
\usepackage{orcidlink}

\usepackage{array}
\newcolumntype{P}[1]{>{\centering\arraybackslash}p{#1}}
\newcolumntype{Q}[1]{>{\raggedleft\arraybackslash}p{#1}}
\newcolumntype{R}[1]{>{\raggedright\arraybackslash}p{#1}}

\newcommand{\rr}{\mathbf{r}}

\newcommand{\kk}{\mathbf{k}}

\newcommand{\GG}{\mathbf{G}}
\newcommand{\gG}{\mathbf{g}}
\newcommand{\D}{\mathrm{d}}

\newcommand{\bvec}[1]{\mathbf{\boldsymbol{#1}}}

\begin{document}

\title{Charge distribution across dislocation networks induced by a strained top layer in hexagonal boron nitride substrates}

\author{Isaac Soltero\,\orcidlink{0000-0002-2593-0891}}
\email{isaac.solteroochoa@manchester.ac.uk}
\affiliation{Department of Physics and Astronomy, University of Manchester, Oxford Road, Manchester, M13 9PL, United Kingdom}
\affiliation{National Graphene Institute, University of Manchester, Booth St.\ E., Manchester, M13 9PL, United Kingdom}
\author{James G. McHugh\,\orcidlink{0000-0001-8509-4883}}
\affiliation{Department of Physics and Astronomy, University of Manchester, Oxford Road, Manchester, M13 9PL, United Kingdom}
\affiliation{National Graphene Institute, University of Manchester, Booth St.\ E., Manchester, M13 9PL, United Kingdom}
\author{Vladimir I. Fal'ko\,\orcidlink{0000-0003-0828-0310}}
\email{vladimir.falko@manchester.ac.uk}
\affiliation{Department of Physics and Astronomy, University of Manchester, Oxford Road, Manchester, M13 9PL, United Kingdom}
\affiliation{National Graphene Institute, University of Manchester, Booth St.\ E., Manchester, M13 9PL, United Kingdom}

\begin{abstract}

Hexagonal boron nitride (hBN) flakes are key building blocks for encapsulating two-dimensional (2D) materials, providing atomically flat surfaces and an excellent dielectric environment for high-mobility field-effect transistors and tunnelling devices. However, strain induced during mechanical exfoliation and assembly of van der Waals heterostructures may lead to plastic deformations of the hBN surface, injecting dislocation lines between the topmost layer and the underlying film. Since a monolayer of hBN is non-centrosymmetric and exhibits a piezoelectric response to deformation, individual dislocations and, in particular their networks, can generate electrostatic potential modulations in the encapsulated 2D material. Here, we examine scenarios in which the top hBN layer is uniaxially strained and/or twisted, and show how lattice reconstruction into dislocation networks leads to the formation of piezoelectric charge hotspots that effectively behave as charged defects.

\end{abstract}

\maketitle

\section{Introduction}

Encapsulation in hexagonal boron nitride (hBN) films is widely used to achieve exceptionally high carrier mobilities and sharp optical resonances in two-dimensional (2D) materials \cite{dean2010boron,Wang2013EdgeContact,Yang2013EpitaxialGraphene,Haigh2012CrossSectional,Purdie2018CleaningInterfaces,withers2015light,withers2015wse2}. Owing to their atomic flatness and large band gap \cite{watanabe2004direct}, hBN films are generally regarded as providing an electrostatically passive environment for such heterostructures \cite{Lee2009InteractionGraphene,Gorbachev2012CoulombDrag,lee2011electron,britnell2012field,mishchenko2014twist}. However, the assembly of van der Waals stacks relies on mechanical exfoliation of hBN flakes, which involves substantial shear forces, while subsequent device fabrication may include deposition of materials with thermal expansion coefficients different from that of hBN \cite{frisenda2018recent}. These processes can introduce strain in the topmost layer of the hBN flake, leading to slight incommensurability with the underlying film or even plastic deformation associated with the injection of dislocations into the bulk.

In this work, we analyse both individual dislocations in hBN and the more complex surface structures that emerge when the top layer is strained, including their dependence on strain magnitude and orientation. Using a stacking-order-dependent interlayer adhesion energy parametrized from density functional theory (DFT), we study the reconstruction of a strained surface layer into perfect 2H stacking domains (vertical alignment of boron and nitrogen atoms in adjacent layers) separated by domain walls, which correspond to in-plane dislocations. We then incorporate the piezoelectric response of the hBN monolayer to inhomogeneous strain and predict the resulting charge distribution across dislocation networks, leading to electrostatic perturbations for charge carriers in encapsulated 2D materials.

\begin{figure}
    \centering
    \includegraphics[width=1.0\linewidth]{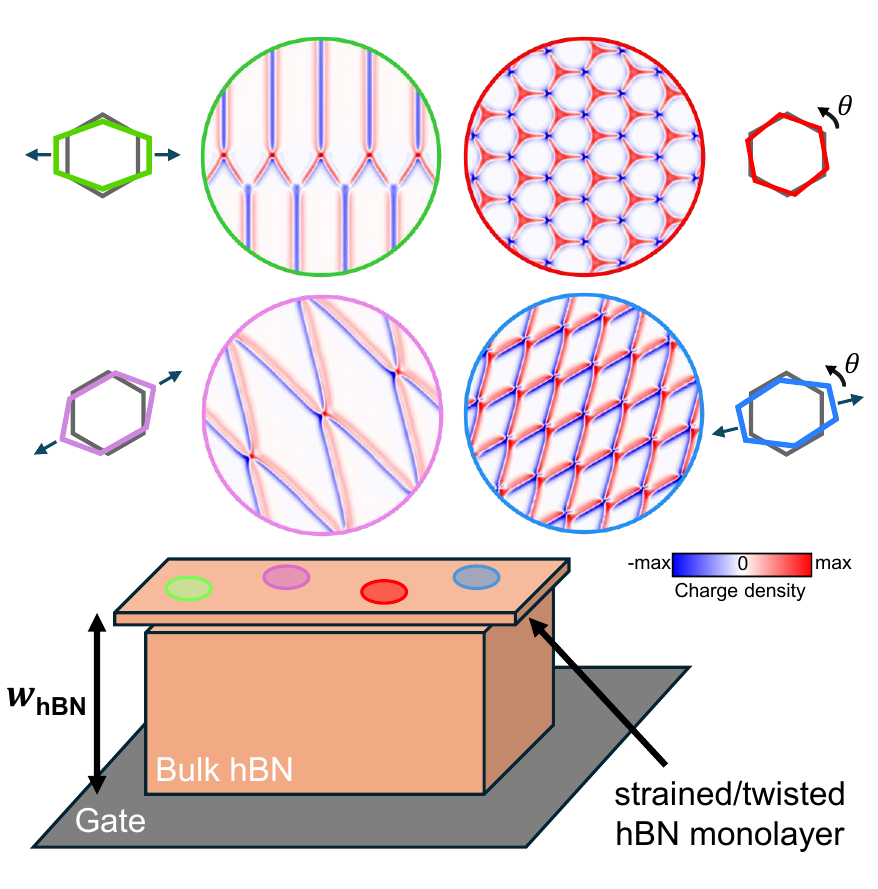}
    \caption{Piezoelectric charge textures from strained and/or twisted monolayer on top of an hBN film with thickness $w_{\rm hBN}$. The presence of inhomogeneous strain on the top layer of hBN gives rise to multiple regions with highly anisotropic piezoelectric charge densities originated from lattice reconstruction effects. Four possible regions are illustrated: two corresponding to uniaxial strain on the top layer (green for strain along the zigzag axis and purple along armchair), one for a relative twist angle (red), and one corresponding to the combination of the two previous cases (blue).}
    \label{fig:Schematics}
\end{figure}

The superlattice formed at the interface between the top monolayer and bulk hBN acts as a magnifying glass for small distortions and relative rotations \cite{cosma2014moire, escudero2024designing}. As a result, dislocation networks formed by lattice reconstruction exhibit significant changes depending on the strain direction and magnitude. In particular, inhomogeneous strain distributions lead to regions with drastically distinct piezoelectric charge densities, as shown schematically in Fig. \ref{fig:Schematics}. We will first analyse the different types of dislocations that can form in these systems and this will guide the analysis of  dislocations in superlattices.

The paper is organized as follows. In Sec. \ref{Sec:Model}, we present the DFT-parameterized stacking-dependent adhesion energy density. Section \ref{Sec:Dislocations} discusses the energetics of interlayer dislocations in hBN multilayers. The geometry and lattice reconstruction of moir{\'e} superlattices formed by a uniaxially strained top layer are analyzed in Sec. \ref{Sec:StrainedTopLayer}, and their electrostatic effects are examined in Sec. \ref{Sec:Piezopotential}. Concluding remarks are given in Sec. \ref{Sec:Conclusions}.

\section{Stacking-dependent adhesion energy}\label{Sec:Model}

The stacking properties of multilayer hBN, in which consecutive layers have antiparallel unit-cell orientations, are modelled using an interpolation formula that captures the energy deviations from the 2H ground-state stacking \cite{enaldiev2020stacking,carr2018relaxation,lebedeva2011interlayer,reguzzoni2012potential,kolmogorov2005registry}. The adhesion energy functional of antiparallel hBN bilayers was computed using \textsc{Quantum ESPRESSO} \cite{Giannozzi2009,Giannozzi2017}. Vanderbilt ultrasoft PBE pseudopotentials were used to approximate core electrons \cite{Garrity2014}, with plane-wave and charge-density cutoffs of $E_{\mathrm{cut}}^{\mathrm{wfc}} = 70$~Ry and $E_{\mathrm{cut}}^{\rho} = 700$~Ry, respectively. Brillouin-zone sampling employed a $31\times31\times1$ Monkhorst--Pack $k$-point mesh with Fermi--Dirac smearing of width $\sigma=0.01$ eV \cite{monkhorst1976special}. Dispersion interactions were treated using the vdW-DF2--c09 functional  \cite{Thonhauser2015PRL,Thonhauser2007PRB,Berland2015RPP,Langreth2009JPCM,Sabatini2012JPCM,Cooper2010PRB}. A $30$~\AA\ vacuum spacing in the out-of-plane direction was used to suppress spurious interactions between periodic images.

Adhesion energies were calculated by implementing an interlayer displacement between hBN monolayers over a uniform two-dimensional grid of lateral translations, where the lateral displacement vector between nitrogen atoms in opposite layers, $\rr_0=(x,y)$, spans the hBN hexagonal unit cell. Atomic positions were relaxed in the out of plane direction for each fixed lateral displacement. The adhesion energy density of each such configuration was extracted as the energy difference between the bilayer stack and two isolated monolayers,
\begin{equation}
    \mathcal{W}(\rr_0) = \frac{E_{\mathrm{BL}}(\rr_0) - 2E_{\mathrm{ML}}}{A},
\end{equation}
where $A=\sqrt{3}a^{2}/2$ is the primitive unit-cell area and $a=0.25$nm the monolayer lattice constant \cite{pease1952x}. $\mathcal{W}(\rr_0)$ was then interpolated over the first three Bragg stars of reciprocal lattice vectors $\GG_{1,2,3}^{(n)}$,
\begin{equation}
\begin{aligned}
    \mathcal{W}(\rr_0) = {} & \sum_{n,m=1}^{3}
    \Big[ w_{n}^{(s)}\cos\!\big(\GG_{m}^{(n)}\!\cdot\!\rr_0\big)
    + w_{n}^{(a)}\sin\!\big(\GG_{m}^{(n)}\!\cdot\!\rr_0\big) \Big],
\end{aligned}
\end{equation}
where the triplet of vectors in each star are related by $120^\circ$ rotation, with magnitudes $|\GG^{(1)}|=4\pi/(\sqrt{3}a)$, $|\GG^{(2)}|=\sqrt{3}\,|\GG^{(1)}|$ and $|\GG^{(3)}|=2|\GG^{(1)}|$.

\begin{table}
    \caption{Parameters of adhesion energy density (eV/nm${}^{2}\times 10^{-3}$) between hBN monolayers.}
    \centering
    \begin{tabular}{c c c c c c}
    \hline
    \hline
         $w_{1}^{(s)}$ & $w_{1}^{(a)}$ & $w_{2}^{(s)}$ & $w_{2}^{(a)}$ & $w_{3}^{(s)}$ & $w_{3}^{(a)}$ \\
    %\hline
         87.91 & 13.66 & -6.18 & 0.0 & -2.67 & -0.60 \\
    \hline
    \hline
    \end{tabular}
    \label{tab:placeholder}
\end{table}

To study multilayer structures, we consider the general case of a dislocation plane or a strained/twisted interface between two hBN films with $N$ and $M$ layers, respectively. The layers in each film are labelled according to their proximity to the dislocation plane (or interface) as $\ell = 1, \dots,N$ and $\ell'=1',\dots,M'$ (see Fig. \ref{fig:DislocationEnergy}(a)). The stacking-dependent adhesion energy density between adjacent layers takes the form
\begin{equation}\label{eq:AdhesionFunc}
\begin{split}
    \mathcal{W}_{k,p}(\rr_0^{(k,p)}) =\, &\sum_{n,m=1}^{3} \Big[ w_{n}^{(s)}\cos(\GG_{m}^{(n)}\cdot\rr_0^{(k,p)})\\
    & + (-1)^{k+1}w_{n}^{(a)}\sin(\GG_{m}^{(n)}\cdot\rr_0^{(k,p)}) \Big],
\end{split}
\end{equation}
where $p=2,\dots,N$ ($p=2',\dots,M'$) and $k=p-1$ for the layers that make up the $N$ ($M$) layer film, and $p=1$ and $k=1'$ for the layers next to the dislocation plane. In Eq. \eqref{eq:AdhesionFunc}, $\rr_0^{(k,p)}$ is the in-plane offset vector between nearest nitrogen atoms of layers $k$ and $p$. Note that the factor $(-1)^{k+1}$ on the second term has been included to account for anti-parallel alignment between consecutive layers.

To model structural deformations around dislocations and strained/twisted interfaces, we introduce the continuous in-plane deformation fields, $\bvec{u}^{(\ell)}$ and $\bvec{u}^{(\ell')}$, such that the offset vector for neighbouring layers within the $N$ or $M$ layer films is
\begin{equation}\label{Eq:StackingBulk}
    \rr_{0}^{(k,p)} = \boldsymbol{\delta}^{({\rm 2H},k)} - \bvec{u}^{(k)} + \bvec{u}^{(p)},
\end{equation}
\begin{equation*}
    \boldsymbol{\delta}^{({\rm 2H},k)}=\frac{a}{2}\bigg(\hat{\bvec{\mathrm{x}}} + (-1)^{k+1}\frac{\hat{\bvec{\mathrm{y}}}}{\sqrt{3}}\bigg),
\end{equation*}
with $\boldsymbol{\delta}^{({\rm 2H},k)}$ the bulk stacking vector. The offset vector at the interface between the $N$ and $M$ layer films, $\rr_{0}^{(1',1)}$, will be specified in Sec. \ref{Sec:Dislocations} and Sec. \ref{Sec:StrainedTopLayer} according to the scenario of interest. The elastic energy cost of the deformation field in each layer is given by
\begin{equation}\label{eq:ElasticFunc}
    \mathcal{B}_{\ell} = \frac{\lambda + \mu}{2}\big(u_{ii}^{(\ell)}\big)^{2} + \frac{\mu}{2} \big[ \big( u_{xx}^{(\ell)} - u_{yy}^{(\ell)} \big)^{2} + 4\big(u_{xy}^{(\ell)}\big)^{2} \big],
\end{equation}
where $u_{ij}^{(\ell)} = (\partial_{i}u_{j}^{(\ell)} + \partial_{j}u_{i}^{(\ell)})/2$ is the 2D strain tensor, $\lambda=627$ eV/nm${}^2$ is the first Lam{\'e} coefficient and $\mu=736$ eV/nm${}^2$ is the shear modulus \cite{androulidakis2018tailoring,enaldiev2024dislocations}.

\section{Dislocations in \lowercase{h}BN}\label{Sec:Dislocations}

Interlayer dislocations separate energetically equivalent 2H stacking regions at the interface between two hBN films. To model such linear defects, we define the in-plane deformation fields along a coordinate axis $\xi$ perpendicular to the dislocation line, which has an angle $\phi$ with respect to the zigzag direction of the hBN monolayers ($x$-axis). We set the boundary conditions as
\begin{subequations}
\begin{equation}
    \bvec{u}^{(\ell)}(-\infty) = \bvec{u}^{(\ell')}(-\infty) = \bvec{0},
\end{equation}
\begin{equation}
     \bvec{u}^{(\ell)}(\infty) -\bvec{u}^{(\ell')}(\infty) =a \hat{\bvec{\mathrm{x}}}.
\end{equation}
\end{subequations}
This defines a dislocation with Burgers vector $\bvec{b} = \bvec{u}^{(\ell)}(\infty) - \bvec{u}^{(\ell')}(\infty)=a\hat{\bvec{\mathrm{x}}}$ (for any $\ell = 1,\dots,N$ and $\ell'=1',\dots,M'$). The local variation of stacking as a function of perpendicular distance to the dislocation line is described by an in-plane offset vector, $\rr_0^{(1',1)}$, as
\begin{equation}
    \rr_{0}^{(1',1)}(\xi) = \boldsymbol{\delta}^{({\rm 2H},1)} - \bvec{u}^{(1')}(\xi)+ \bvec{u}^{(1)}(\xi) .
\end{equation}

\begin{figure}
    \centering
    \includegraphics[width=0.9\linewidth]{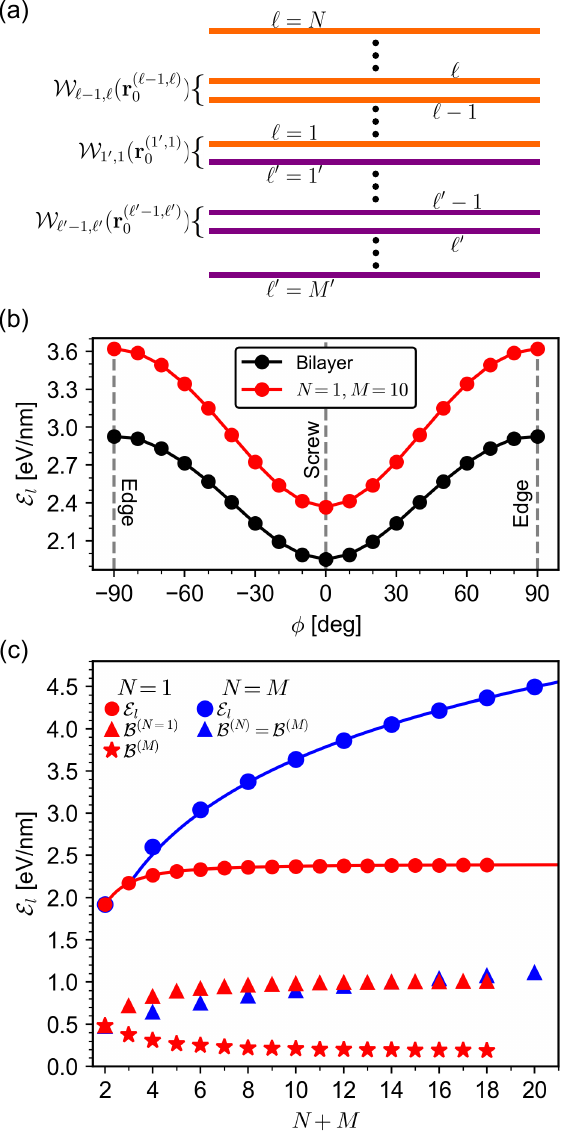}
    \caption{(a) Diagram illustrating a dislocation or strained/twisted interface between two hBN films with $N$ and $M$ layers, indicated by orange and blue colours, respectively. The labelling of monolayers according to their proximity to the interface and the adhesion between each pair of layers is shown. (b) Orientation dependence of the dislocation energy per unit of length (defined in Eq. \eqref{Eq:DislocationEnergyDensity}), described in terms of the angle $\phi$ between the dislocation axis and the Burgers vector. Black dots indicate the case of a dislocation in a bilayer ($N=1$, $M=1$), and red dots a dislocation between a monolayer and a multilayer film. (c) Screw dislocation ($\phi=0^{\circ}$) energy per unit of length  as a function of the total number of layers $N+M$. Blue dots correspond to structures with equal number of layers in each side of the dislocation plane, $N=M$, and red dots correspond to a dislocation between a monolayer ($N=1$) and an $M$ layer film. Curves represent the power law and logarithmic fittings defined in Eqs. \eqref{Eq:LogFit} and \eqref{Eq:PowerFit}, respectively. The total elastic energy $\mathcal{B}$ of the constituent films is indicated by triangles and stars.}
    \label{fig:DislocationEnergy}
\end{figure}

The energy of the structure is defined by the functional
\begin{equation}\label{Eq:DislocationEnergy}
\begin{split}
    \mathcal{E} = \int_{-\infty}^{\infty} \D \xi \, \Bigg[ &\mathcal{W}_{1',1}(\rr_{0}^{(1',1)}) + \mathcal{B}_{1} + \mathcal{B}_{1'}\\
    &+ \sum_{\ell=2}^{N}\Big\{\mathcal{W}_{\ell-1,\ell}(\rr_{0}^{(\ell-1,\ell)}) + \mathcal{B}_{\ell} \Big\} \\
    &+ \sum_{\ell'=2'}^{M'}\Big\{\mathcal{W}_{\ell'-1,\ell'}(\rr_{0}^{(\ell'-1,\ell')}) + \mathcal{B}_{\ell'} \Big\} \Bigg].
\end{split}
\end{equation}
Due to the contribution of the infinite 2H stacking regions, the energy density \eqref{Eq:DislocationEnergy} diverges. Therefore, the optimal configuration for the deformation fields $\bvec{u}^{(\ell)}$ and $\bvec{u}^{(\ell')}$ is found by minimising the dislocation energy per unit of length,
\begin{equation}\label{Eq:DislocationEnergyDensity}
\begin{split}
    \mathcal{E}_{l} = \mathcal{E} - \int_{-\infty}^{\infty} \D\xi \Bigg[ &\mathcal{W}_{1',1}(\boldsymbol{\delta}^{({\rm 2H},1)}) + \sum_{\ell=2}^{N} \mathcal{W}_{\ell-1,\ell}(\boldsymbol{\delta}^{({\rm 2H},\ell)}) \\
    & + \sum_{\ell'=2'}^{M'} \mathcal{W}_{\ell'-1,\ell'}(\boldsymbol{\delta}^{({\rm 2H},\ell')}) \Bigg],
\end{split}
\end{equation}
where the second term subtracts the energy of a multilayer crystal with unperturbed 2H stacking. This minimisation is performed by progressively increasing the length of the $\xi$ axis and reducing the step until we achieve convergence of the value of $\mathcal{E}_{l}$. In line with previous studies on dislocations in graphene and transition metal dichalcogenide multilayers \cite{lebedeva2016dislocations,enaldiev2020stacking,enaldiev2024dislocations,soltero2025interlayer}, the optimal orientation of dislocations in bilayer hBN ($N=1$, $M=1$) corresponds to $\phi=0^{\circ}$ (see Fig. \ref{fig:DislocationEnergy}(b)), where the Burgers vector along the zigzag direction has parallel orientation to the dislocation line ($\xi =y$, screw dislocation). The most energetically unfavourable case corresponds to a dislocation axis perpendicular to the Burgers vector ($\xi=x$, edge dislocation), with an energy difference of $\sim 1$eV/nm.

The evolution of the screw dislocation energy density as a function of the total number of layers, $N+M$, shows a contrasting behaviour for symmetrical structures with equal number of layers on both sides of the dislocation plane ($N=M$) with respect to dislocations between a monolayer ($N=1$) and a multilayer film (see Fig. \ref{fig:DislocationEnergy}(c)). For symmetric structures, the dislocation energy follows a logarithmic growth as the thickness is increased,
\begin{equation}\label{Eq:LogFit}
    \mathcal{E}_{l}^{(N=M)}(N+M) = \eta \ln\big[ \nu (N+M) \big],
\end{equation}
where $\eta=1.22$ eV/nm and $\nu=1.86$. On the other hand, energy of screw dislocations between monolayers ($N=1$) and $M$ layer films increases as a power law,
\begin{equation}\label{Eq:PowerFit}
    \mathcal{E}_{l}^{(N=1)}(N+M) = \mathcal{E}_{l}^{(N=1)}(\infty) - \frac{\beta}{(N+M)^{\gamma}},
\end{equation}
with the fitting parameters $\beta=1.87$ eV/nm, $\gamma = 1.94$, and the saturation value $\mathcal{E}_{l}^{(N=1)}(\infty) = 2.39$ eV/nm. To understand this qualitatively different behaviour between structures with same number of layers, we calculate the total elastic energy accumulated in the films on both sides of the dislocation, $\mathcal{B}^{(N/M)}\equiv \int_{-\infty}^{\infty} \D y \sum_{\ell/\ell'} \mathcal{B}_{\ell/\ell'}$. While for symmetrical structures ($N=M$) elastic energy energy is distributed equally on both films, for asymmetrical structures the elastic energy tends to accumulate in the thinnest film, reflecting the lower rigidity (see Fig. \ref{fig:DislocationEnergy}(c)).

\section{Reconstructed moir{\'e} superlattices from uniaxially strained top layer}\label{Sec:StrainedTopLayer}

Building on our analysis of single dislocations in multilayer systems, we now consider a uniaxially strained monolayer stacked on an undeformed film. For small values of uniaxial strain and relative twist, the resulting moir{\'e} superlattice undergoes substantial lattice reconstruction, leading to the formation of large stacking domains separated by a network of highly strained dislocations. The character of these dislocations is determined entirely by the geometry of the moir{\'e} pattern. Below, we identify the dislocation types in terms of the applied uniaxial strain and the relative twist angle of the upper hBN layer. To model the interface between a monolayer and bulk hBN, we consider a strained and/or twisted interface between structures with $N=1$ and $M=8$ layers. For this configuration, the strain fields in both films are converged, and no significant changes are expected upon further increasing $M$.

We consider the lattice vectors of the monolayers forming bulk hBN as $\bvec{a}_{1,b}=a\hat{\bvec{\mathrm{x}}}$ and $\bvec{a}_{2,b}=\frac{a}{2}(\hat{\bvec{\mathrm{x}}} + \sqrt{3}\hat{\bvec{\mathrm{y}}})$, together with the reciprocal lattice vectors $\bvec{G}_{1,b} = \frac{2\pi}{a} ( \hat{\bvec{\mathrm{x}}} - \frac{1}{\sqrt{3}}\hat{\bvec{\mathrm{y}}} )$ and $\bvec{G}_{2,b} = \frac{4\pi}{\sqrt{3}a}\hat{\bvec{\mathrm{y}}}$. Following Refs. \onlinecite{escudero2024designing,escudero2025geometrical,hermann2012periodic,cosma2014moire}, uniaxial strain is applied on the top layer of bulk hBN along an axis with angle $\alpha$ with respect to the $x$-axis (zigzag direction), such that the strain tensor reads as
\begin{equation}\label{Eq:StrainTensor}
\begin{split}
    \mathcal{U} =&\, \mathcal{R}(\alpha)
    \begin{pmatrix}
        \epsilon & 0 \\ 0 & -\sigma\epsilon
    \end{pmatrix}
    \mathcal{R}(-\alpha)\\
    =&\, \epsilon \begin{pmatrix}
        \cos^2\alpha - \sigma\sin^2\alpha & (1+\sigma)\sin\alpha\cos\alpha \\
        (1+\sigma)\sin\alpha\cos\alpha & \sin^2\alpha - \sigma\cos^2\alpha
    \end{pmatrix},
\end{split}
\end{equation}
where $\mathcal{R}(\alpha)$ is the 2D rotation operator, $\epsilon$ is the magnitude of strain, and $\sigma=0.21$ is the Poisson ratio for hBN \cite{androulidakis2018tailoring}. We also include a twist of the strained monolayer by an angle $\theta\ll1$, resulting in the following lattice vectors for the top layer:
\begin{equation}
    \bvec{a}_{i,t} = (\mathbbm{I}+\mathcal{U})\mathcal{R}(\theta)\bvec{a}_{i,b}, \qquad i=1,2.
\end{equation}
From the definition $\bvec{a}_{i,t}\cdot\bvec{G}_{j,t}=2\pi\delta_{ij}$, the corresponding reciprocal lattice vectors are
\begin{equation}
\begin{split}
    \GG_{i,t} =&\, (\mathbbm{I}+\mathcal{U})^{-1} \mathcal{R}(\theta) \GG_{i,b}\\
    \approx&\, (\mathbbm{I}-\mathcal{U}) \mathcal{R}(\theta) \GG_{i,b},
\end{split}
\end{equation}
where we have taken into account $\epsilon\ll1$. The moir{\'e} reciprocal lattice vectors are
\begin{equation}\label{Eq:moireReciprocal}
\begin{split}
    \gG_{i} =&\, \GG_{i,t} - \GG_{i,b}\\
    \approx&\, \big[ \mathcal{R}(\theta) - \mathcal{U} - \mathbbm{I} \big] \GG_{i,b} \\
    =& \, \theta \hat{\bvec{\mathrm{z}}}\times \GG_{i,b} - \mathcal{U}\GG_{i,b}. 
\end{split}
\end{equation}
Hence, we can write the moir{\'e} lattice vectors as
\begin{equation}\label{eq:moirevectors}
    \bvec{A}_{i} \approx \big[ \{ \mathcal{R}(\theta) -\mathcal{U} - \mathbbm{I}\}^{-1} \big]^{\dagger} \bvec{a}_{i,b}.
\end{equation}
In particular, for $\theta=0^{\circ}$, the superlattice vectors are
\begin{subequations}\label{eq:moirevectors0}
\begin{equation}
    \bvec{A}_{1} =\, \frac{a}{\sigma\epsilon}\begin{pmatrix}
        \sin^2\alpha - \sigma\cos^2\alpha \\ -(1+\sigma)\sin\alpha\cos\alpha
    \end{pmatrix},
\end{equation}
\begin{equation}
    \bvec{A}_{2} = \frac{a}{2\sigma\epsilon}\begin{pmatrix}
        \sin^2\alpha - \sigma\cos^2\alpha - \sqrt{3}(1+\sigma)\sin\alpha\cos\alpha \\
        -(1+\sigma)\sin\alpha\cos\alpha + \sqrt{3}(\cos^2\alpha - \sigma\sin^2\alpha)
    \end{pmatrix}.
\end{equation}
\end{subequations}
A special scenario is the case when the moir{\'e} lattice vectors become collinear, which results in a quasi-one-dimensional superlattice \cite{cosma2014moire,sinner2023strain,escudero2024designing}. This corresponds to the collapse of the moir{\'e} Brillouin zone when ${\rm det}[\mathcal{R}(\theta) - \mathcal{U} - \mathbbm{I}]=0$. For $\theta\ll 1$ this reads as
\begin{equation}
    \frac{\theta_{\rm 1D}}{\epsilon} \approx \pm \sqrt{\sigma} \approx \pm 0.46,
\end{equation}
independently of the orientation of the axes of the top layer strain tensor.

\begin{figure*}
    \centering
    \includegraphics[width=\linewidth]{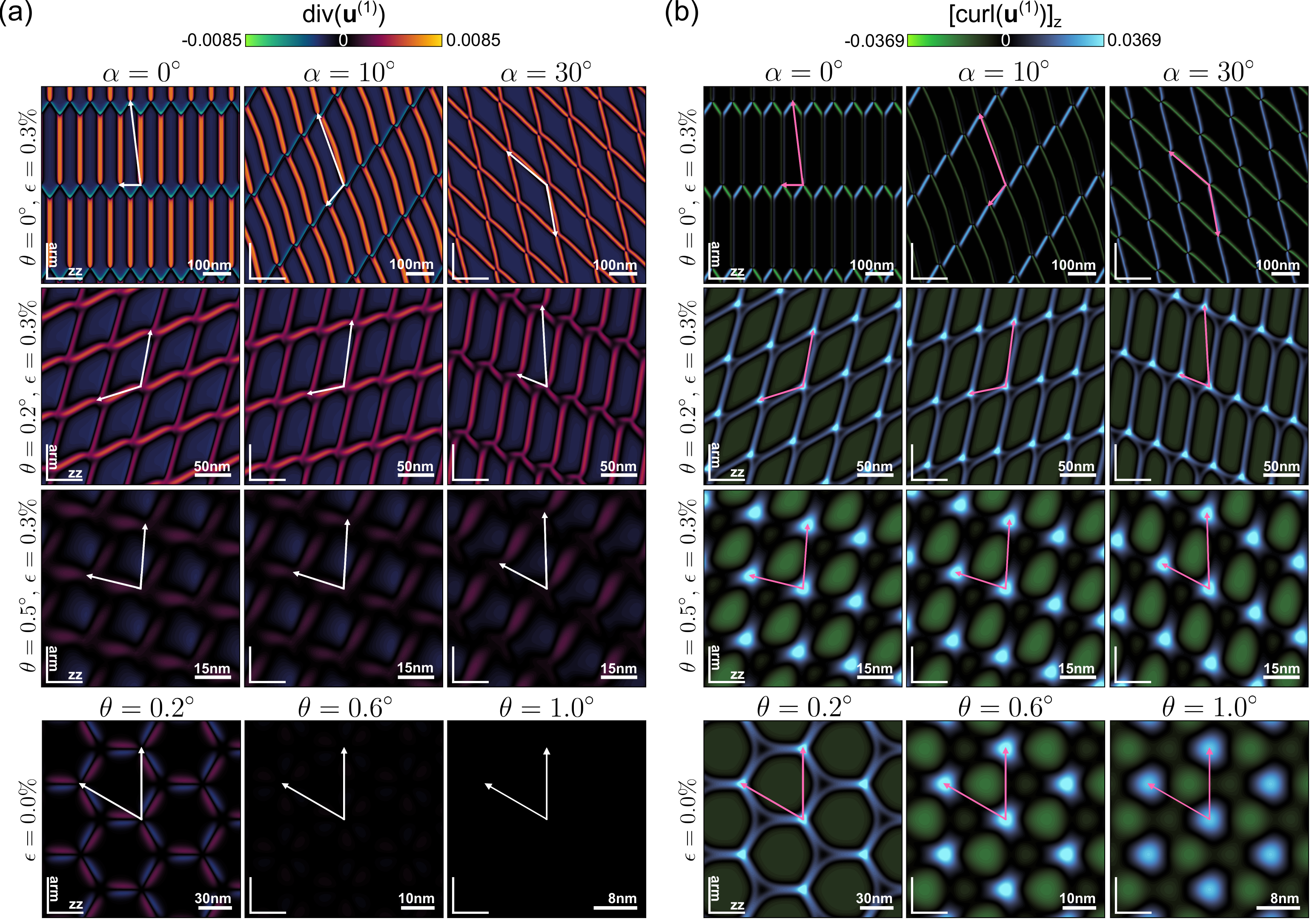}
    \caption{In-plane displacement field maps of reconstructed moir{\'e} superlattices. (a) Divergence and (b) curl maps for displacement fields in the top layer ($\bvec{u}^{(1)}$). The structure corresponds to a uniaxially strained monolayer along an axis with angle $\alpha$ with respect to the zigzag direction stacked on top of bulk hBN with a relative twist angle $\theta$. Each row in the top panel corresponds to a different configuration of strain $\epsilon$ and twist angle $\theta$, whereas columns are the different axes where uniaxial strain is applied, with $\alpha=0^{\circ}$ being the zigzag direction and $\alpha=30^{\circ}$ the armchair direction. The last row corresponds to a purely twisted (unstrained) top layer.}
    \label{fig:Divergence_Curl}
\end{figure*}

The optimal configuration for the in-plane displacement fields in the moir{\'e} superlattice, $\bvec{u}^{(1)}$ and $\bvec{u}^{(\ell')}$, is found by minimising the energy of a single supercell, taking into account both elastic and adhesion contributions (see Appendix \ref{Append:LatticeRelaxation}). For this, we parametrise the nitrogen-to-nitrogen offset vector at the interface between the monolayer and the $M$ layer film as
\begin{equation}\label{Eq:StackingStrain}
\begin{split}
    \rr_{0}^{(1',1)}(\rr) =&\, \big[ (\mathbbm{I}+\mathcal{U})\mathcal{R}(\theta) - \mathbbm{I} \big]\rr  - \bvec{u}^{(1')}(\rr)  + \bvec{u}^{(1)}(\rr)\\
    %\approx&\, \big[ \mathcal{R}(\theta) + \mathcal{U} - \mathbbm{I} \big]\rr - \bvec{u}^{(1')}(\rr) + \bvec{u}^{(1)}(\rr)\\
    \approx&\, \theta \hat{\bvec{\mathrm{z}}}\times\rr + \mathcal{U}\rr - \bvec{u}^{(1')}(\rr) + \bvec{u}^{(1)}(\rr) .
\end{split}
\end{equation}
Representative examples of lattice reconstruction are shown in Fig. \ref{fig:Divergence_Curl}, which displays maps of the divergence and curl of the displacement field in the top layer, corresponding to the hydrostatic (Fig. \ref{fig:Divergence_Curl}(a)) and shear strain (Fig. \ref{fig:Divergence_Curl}(b)) components, respectively, for $\epsilon=0.3\%$ and various strain directions and twist angles. Each column corresponds to strain applied along different crystallographic axes, ranging from the zigzag ($\alpha=0^{\circ}$) to armchair direction ($\alpha=30^{\circ}$).

For strain along the zigzag direction ($\alpha=0^{\circ}$) at $\theta=0^{\circ}$ (see top left panel in Fig. \ref{fig:Divergence_Curl}(a,b)), strongly distorted hexagonal domains form, separated by two types of dislocations connecting the nitrogen-on-nitrogen and boron-on-boron stacking regions. The first type consists of long edge dislocations extending to the farthest node, characterized by predominantly hydrostatic strain and negligible shear (${\rm div}(\bvec{u}^{(1)})\neq 0$, ${\rm curl}(\bvec{u}^{(1)})\approx 0$). The second type comprises shorter dislocations connecting the two nearest nodes; these have mixed (edge and screw) character and exhibit both hydrostatic and shear components (${\rm div}(\bvec{u}^{(1)})\neq 0$, ${\rm curl}(\bvec{u}^{(1)})\neq 0$).

In contrast, for strain along the armchair direction ($\alpha=30^{\circ}$, see top right panel in Fig. \ref{fig:Divergence_Curl}(a,b)), reconstruction produces rhombic domains due to the merging of nitrogen-on-nitrogen and boron-on-boron stacking regions. The resulting network contains a single type of dislocation with both hydrostatic and shear components. Upon increasing the twist angle into the regime $\theta[{\rm rad}]>\epsilon$ ($\theta>0.17^{\circ}$ for $\epsilon =0.3\%$), dislocations progressively evolve toward purely screw character, consistent with the network observed in strain-free, marginally twisted interfaces (see bottom row in Fig. \ref{fig:Divergence_Curl}).

\section{Strain-induced piezoelectric potentials}\label{Sec:Piezopotential}

\begin{figure}
    \centering
    \includegraphics[width=0.9\linewidth]{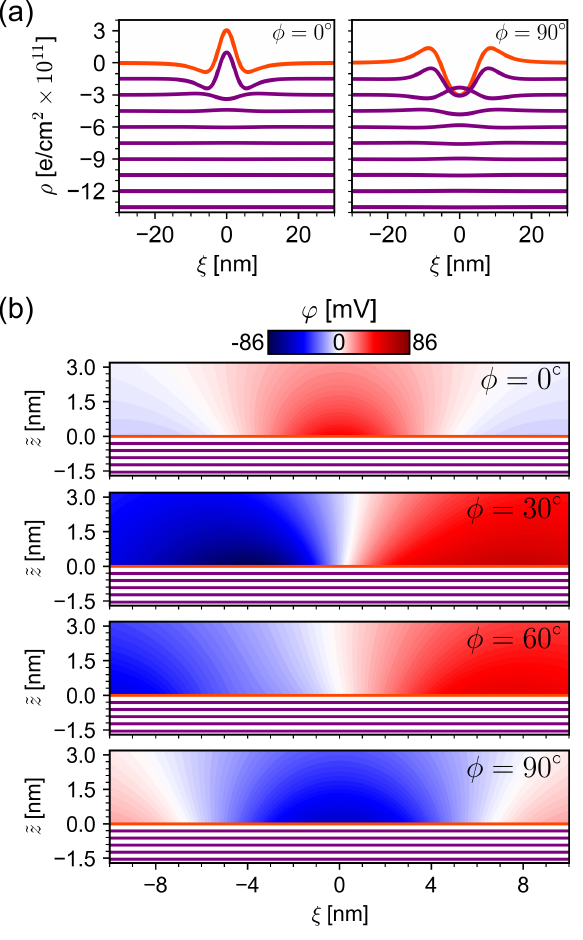}
    \caption{Electrical properties of dislocations between a monolayer and bulk hBN. (a) Piezoelectric charge profiles in each layer across a screw ($\phi=0^{\circ}$, left panel) and an edge dislocation (($\phi=90^{\circ}$, right panel)) in hBN films  with $N=1$ and $M=10$, indicated by orange and purple, respectively. (b) Piezoelectric potential maps as a function of the distance to the hBN surface, $\tilde{z}=z-w_{\rm hBN}$. We show maps for screw, edge, and two mixed ($\phi=30^{\circ},60^{\circ}$) dislocations.}
    \label{fig:Potential_Dislocation}
\end{figure}

Due to the non-centrosymmetric nature of monolayer hBN crystals, the inhomogeneous distribution of strain around dislocations results in a piezoelectric charge profile \cite{enaldiev2020stacking},
\begin{equation}\label{Eq:Piezocharge}
    \rho^{(\ell)} = -e_{11}^{(\ell)}\Big[ 2\partial_{x}u_{xy}^{(\ell)} + \partial_{y}\big( u_{xx}^{(\ell)} - u_{yy}^{(\ell)} \big) \Big],
\end{equation}
where the piezocoefficient alternates sign for neighbouring layers, $e_{11}^{(\ell)}=(-1)^{\ell}e_{11}$, with $e_{11} = 3.71\times 10^{-10}$C/m \cite{duerloo2012intrinsic}. We take $e_{11}^{\ell}>0$ for monolayers where the boron to nitrogen bond direction coincides with the positive $y$-direction, and $e_{11}^{\ell}<0$ for the opposite case. In Fig. \ref{fig:Potential_Dislocation}(a) we show the piezoelectric charge distribution around a screw ($\phi=0^{\circ}$) and an edge dislocation ($\phi=90^{\circ}$). In both cases, the piezoelectric charge density is confined to the first four layers near the dislocation. Thus, the convergence of energy and strain discussed in Sec. \ref{Sec:Dislocations} also implies convergence of the charge distribution.

To analyse the electrostatic potential produced by piezoelectric charges, we consider a general setup where the region $0\leq z\leq w_{\rm hBN}$ represents the hBN substrate of thickness $w_{\rm hBN}=50$nm, with a dielectric tensor ${\rm diag}(\varepsilon_{\parallel}^{(1)},\varepsilon_{\parallel}^{(1)},\varepsilon_{\perp}^{(1)})$ ($\varepsilon_{\parallel}^{(1)}=6.9$ and $\varepsilon_{\parallel}^{(1)}=3.0$ \cite{laturia2018dielectric,ferreira2022scaleability}). The region above the substrate is occupied by an arbitrary medium with dielectric tensor ${\rm diag}(\varepsilon_{\parallel}^{(2)},\varepsilon_{\parallel}^{(2)},\varepsilon_{\perp}^{(2)})$. To model a realistic experimental scenario, we also include a gate at $z=0$.

The electric potential due to the charge density localized in the different hBN layers is obtained by solving the Poisson equation,
\begin{equation}\label{Eq:PoissonEq}
\begin{split}
    \varepsilon_{\parallel}(z)\nabla_{\rr}^2 \varphi(\rr,z) + \partial_{z}\big[&\varepsilon_{\perp}(z) \partial_{z}\varphi(\rr,z)\big]\\
    &= -4\pi\sum_{p=\ell,\ell'}\rho^{(p)}(\rr)\delta (z-z_p),
\end{split}
\end{equation}
\begin{equation*}
    \varepsilon(z) = \begin{cases}
        {\rm diag}(\varepsilon_{\parallel}^{(1)},\varepsilon_{\parallel}^{(1)},\varepsilon_{\perp}^{(1)}), & 0< z\leq w_{\rm hBN},  \\
        {\rm diag}(\varepsilon_{\parallel}^{(2)},\varepsilon_{\parallel}^{(2)},\varepsilon_{\perp}^{(2)}), & w_{\rm hBN}<z.
    \end{cases}
\end{equation*}
For this, we introduce the Green's function, $G(\kk;z,z')$, for the Fourier transformed Eq. \eqref{Eq:PoissonEq} over the in-plane coordinates and point source $0<z'\leq w_{\rm hBN}$,
\begin{equation}
    \partial_z\big[\varepsilon_{\perp}(z)\partial_{z} G(\kk ;z,z')\big] - \varepsilon_{\parallel}(z) |\kk|^{2} G(\kk ;z,z') = \delta(z-z'),
\end{equation}
with boundary conditions
\begin{subequations}
\begin{equation}
    \varepsilon_{\perp}^{(2)}\partial_{z}G(\kk;w_{\rm hBN}^{+},z') - \varepsilon_{\perp}^{(1)}\partial_{z}G(\kk;w_{\rm hBN}^{-},z') = 0,
\end{equation}
\begin{equation}
    \varepsilon_{\perp}^{(1)}\Big[\partial_{z}G(\kk;(z')^{+},z') - \partial_{z}G(\kk;(z')^{-},z')\Big] = 1,
\end{equation}
\begin{equation}
    G(\kk;0,z') = 0.
\end{equation}
\end{subequations}

The resulting piezopotentials in the region above the hBN substrate ($z>w_{\rm hBN}$) are
\begin{equation}\label{Eq:PotentialGreen}
    \varphi(\rr,z) = -4\pi\int \frac{\D^{2}k}{(2\pi)^{2}}\, e^{i\kk\cdot\rr}\sum_{p=\ell,\ell'}G(\kk;z,z_{p})\tilde{\rho}^{(p)}(\kk),
\end{equation}
with
\begin{equation}
\begin{split}
    G(\kk;z,z') =&\,\frac{2\sinh(|\kk|\tilde{\kappa}_{1}z')}{[\tilde{\varepsilon}_{2} - \tilde{\varepsilon}_{1} - e^{2|\kk|\tilde{\kappa}_{1}w_{\rm hBN}}(\tilde{\varepsilon}_{1} + \tilde{\varepsilon}_{2})]|\kk|}\\
    &\times {\rm exp}\Big\{-|\kk|\big[\tilde{\kappa}_{2}z - (\tilde{\kappa}_{1} + \tilde{\kappa}_{2})w_{\rm hBN}\big]\Big\}.
\end{split}
\end{equation}
Here, we have defined
\begin{equation}
    \tilde{\rho}^{(p)}(\kk) = \int \D^{2}r\, e^{-i\kk\cdot\rr}\rho^{(p)}(\rr),
\end{equation}
and the effective dielectric constants $\tilde{\varepsilon}_{i}=\sqrt{\varepsilon_{\perp}^{(i)}\varepsilon_{\parallel}^{(i)}}$ and $\tilde{\kappa}_{i}=\sqrt{\varepsilon_{\parallel}^{(i)}/\varepsilon_{\perp}^{(i)}}$.

Before presenting the results for dislocation networks below, we first analyse the electrostatic potential generated by individual dislocations at the interface between a monolayer and bulk hBN (see Fig. \ref{fig:Potential_Dislocation}(b)), assuming vacuum in the upper region ($\varepsilon_{\parallel}^{(2)} = \varepsilon_{\perp}^{(2)}=1$). In screw ($\phi=0^{\circ}$) and edge dislocations ($\phi=90^{\circ}$), the potential is symmetric, with a peak value at the centre of the dislocation line. In contrast, mixed dislocations ($\phi=30^{\circ}$ and $\phi=60^{\circ}$) produce a potential of opposite sign on either side of the dislocation line, changing sign at its centre.

\begin{figure*}
    \centering
    \includegraphics[width=1.0\linewidth]{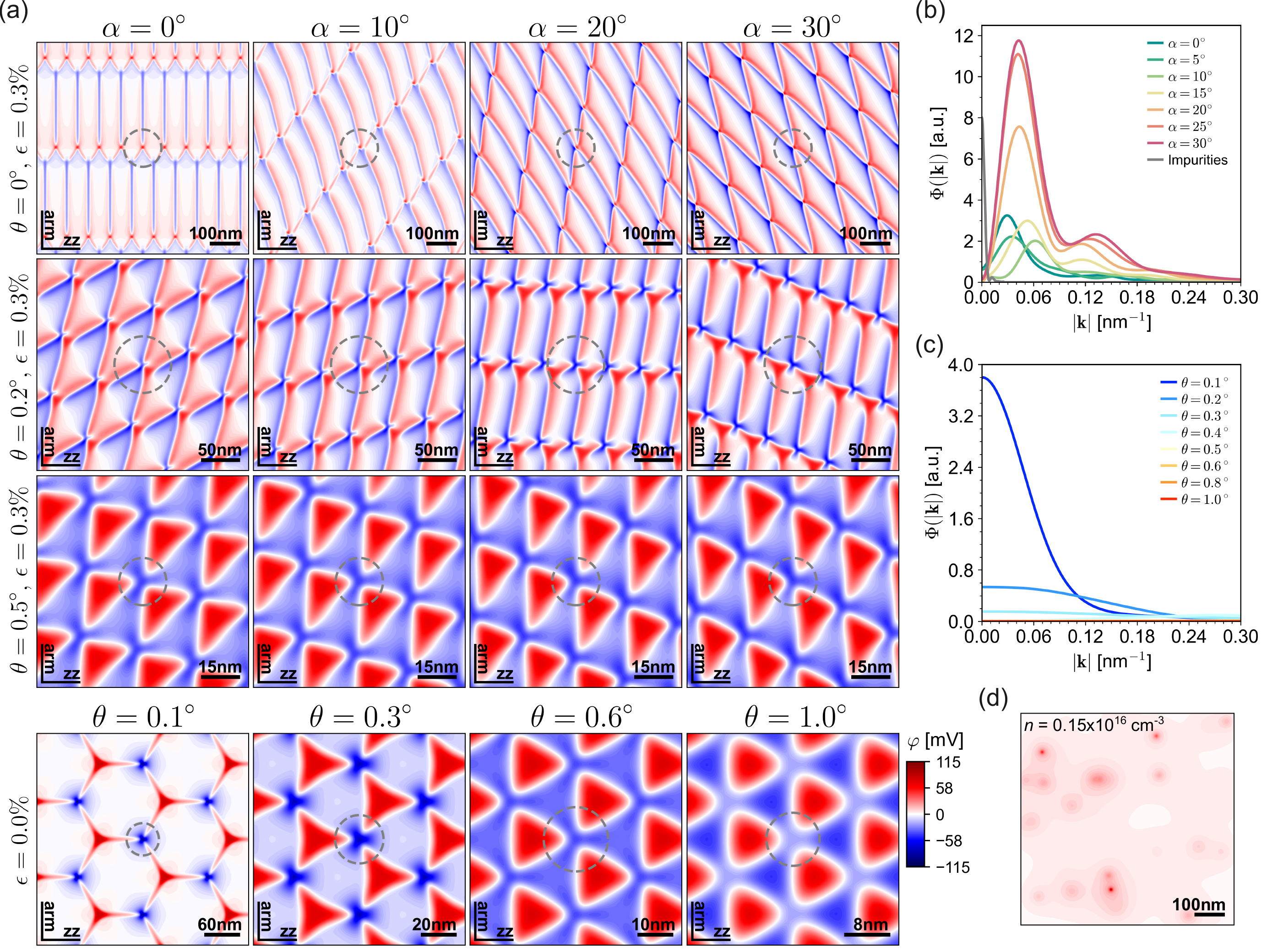}
    \caption{ (a) Piezopotential maps 0.3nm above a strained and/or twisted monolayer stacked on bulk hBN. Upper panel corresponds to a $\epsilon=0.3\%$ uniaxially strained monolayer, where each column is a different direction for the applied strain ($\alpha=0^{\circ}$ for zigzag and $\alpha=30^{\circ}$ for armchair) and rows are different twist angles. Bottom panel is the twist angle dependence for an unstrained monolayer. (b) Form factor, defined as the angular average of the Fourier transform of the piezopotential peaks, for purely strained upper monolayer ($\theta=0^{\circ}$), and for (c) twisted upper monolayer without strain ($\epsilon=0\%$). (d) Potential produced by an arbitrary distribution of charged point-defects with density $n=0.15\times10^{16}$cm$^{-3}$ in a 50nm thick film.}
    \label{fig:Potential_Uniaxial}
\end{figure*}

To calculate the reconstruction-induced piezoelectric potential in dislocation networks, we take into account that the piezoelectric charge texture (Eq. \eqref{Eq:Piezocharge}) has the same periodicity as the moir{\'e} superlattice, which allows to represent each of the layer charge densities in the form of a Fourier series,
\begin{equation}\label{Eq:ChargeFourier}
    \rho^{(\ell)}(\rr) = \sum_{n,m=0}^{\infty}\Big[ \rho_{n,s}^{(\ell)}\cos(\bvec{g}_{nm}\cdot \rr) + \rho_{n,a}^{(\ell)}\sin(\bvec{g}_{nm}\cdot \rr) \Big],
\end{equation}
with $\gG_{nm}=n\gG_{1} + m\gG_{2}$ moir{\'e} reciprocal lattice vectors. The piezopotential above the uniaxially strained monolayer is calculated by substituting \eqref{Eq:ChargeFourier} in Eq. \eqref{Eq:PotentialGreen},
\begin{equation}
\begin{split}
    \varphi(\rr,z) = -4\pi \sum_{p=\ell,\ell'}\sum_{n,m=0}^{\infty}\Big[& G(\bvec{g}_{nm};z,z_{p})\big\{  \rho_{n,s}^{(p)}\cos(\bvec{g}_{nm}\cdot \rr) \\
    &+ 
    \rho_{n,a}^{(p)}\sin(\bvec{g}_{nm}\cdot \rr) \big\}\Big].
\end{split}
\end{equation}
To consider a typical device configuration where a material is placed on an hBN substrate and encapsulated by a second hBN layer, we assume the upper medium to be hBN as well ($\varepsilon_{\parallel}^{(2)} = 6.9, \varepsilon_{\perp}^{(2)}=3.0$). Fig. \ref{fig:Potential_Uniaxial}(a) shows the resulting potential profiles 0.3nm above the interface (the graphene/hBN interlayer distance) for a $\epsilon=0.3\%$ uniaxially strained layer (top panel) and a twisted monolayer (bottom panel). In each case, we can infer the dislocation types from the potential profiles.

For a purely strained monolayer ($\theta=0^{\circ}$, see top row in Fig. \ref{fig:Potential_Uniaxial}(a)), long edge dislocations appear only when the top monolayer is strained along the zigzag direction ($\alpha=0^{\circ}$), which is indicated by the symmetric potential around the dislocations. When strain is applied along other axes, the dislocations exhibit a mixed character, with an asymmetric potential relative to their centre. As the twist angle increases to the regime $\theta[{\rm rad}]\approx \epsilon$ (middle row in top panel of Fig. \ref{fig:Potential_Uniaxial}(a)), the network consists of mixed dislocations, where the relative orientation between them depends strongly on the direction of strain. For twist angles $\theta[{\rm rad}]>\epsilon$ (third row in top panel of Fig. \ref{fig:Potential_Uniaxial}(a)), the piezopotential profiles become independent of the strain direction. In contrast, for a twisted monolayer without strain (bottom panel in Fig. \ref{fig:Potential_Uniaxial}(a)), screw dislocations appear only for twist angles $\theta<0.3^{\circ}$.

For all strain directions considered, ranging from zigzag to armchair, we observe the formation of peaks in the potential magnitude at the network nodes (AA stacking). These peaks act as scattering centres for charge carriers in the material placed on top of the hBN substrate. The majority potential sign of these scattering centres changes gradually with respect to the axis of strain applied to the top layer. For strains predominantly along the zigzag direction ($\alpha<15^{\circ}$), the peaks are majority positive, while for strains along the armchair direction ($\alpha>15^{\circ}$), the dominant potential is negative. Notably, at the limit $\alpha=0^{\circ}$, the potential at the node is purely positive. However, at $\alpha=30^{\circ}$, although the potential is primarily negative, there remains a positive potential peak. For an unstrained monolayer with twist angle, a similar accumulation of potential at the nodes is observed, however, the potential is exclusively negative and decays rapidly with increasing the misalignment.

To quantify the momentum transfer induced by the calculated potential profile centres leading to scattering of charge carriers in the hBN encapssulated material, we take the Fourier transform of the potential on a disk around the dislocation node (indicated by dashed circles in Fig. \ref{fig:Potential_Uniaxial}(a)), $\tilde{\varphi}(\kk)$, and define the following form factor as the angular average of its magnitude:
\begin{equation}
    \Phi(|\kk|) \equiv \frac{1}{2\pi}\int_{0}^{2\pi} \D\vartheta_{\kk}\, |\tilde{\varphi}(\kk)|^{2}.
\end{equation}
The strain-direction dependence of this form factor for non-twisted surface layers, shown in Fig. \ref{fig:Potential_Uniaxial}(b) for $\epsilon=0.3\%$, exhibits a single peak at $|\kk|\sim$ 0.03nm$^{-1}$ for $\alpha=0^{\circ}$. As the strain orientation approaches $\alpha=30^{\circ}$, the spectrum evolves smoothly into a two-peak structure at $|\kk|\sim$ 0.045nm$^{-1}$ and $\sim$0.14nm$^{-1}$. The emergence of one or two characteristic wave vectors reflects the dominant piezopotential contribution at the nodes of the dislocation network. As the strain is increased within $\epsilon<1\%$, the form factor peak shows a blueshift towards 0.1nm$^{-1}$ accompanied by a reduction in amplitude and increased broadening (see Appendix \ref{Append:StrainTwistPotential}). Furthermore, potential build-up at the nodes is restricted to small misalignment of the top layer (Fig. \ref{fig:Potential_Uniaxial}(c)), with a $|\kk|=0$ peak in the form factor only for $\theta<0.3^{\circ}$.

For comparison, we calculate the electrostatic potential generated by an arbitrary distribution of point charges within the hBN film. We find that a charge density $n\approx0.15\times10^{16}$cm${}^{-3}$ produces a surface potential comparable to that induced by the dislocation networks (Fig. \ref{fig:Potential_Uniaxial}(d)).

\section{Conclusions}\label{Sec:Conclusions}

Overall, we offer a detailed analysis of the structural and electrical properties of dislocation networks formed by a top strained monolayer on an hBN film. By studying the strain fields and electrical potentials produced by individual dislocations, we identify the character of dislocation networks in the highly anisotropic moir{\'e} patterns arising from uniaxial strain. For configurations where the upper monolayer is strained by $\epsilon$ and twist angle $\theta[{\rm rad}]\lesssim\epsilon$, long dislocation networks are formed, with strong dependence on the direction of the applied strain. For larger misalignments ($\theta[{\rm rad}]>\epsilon$), the reconstruction patterns become independent of both the strain magnitude and direction, resembling those found in twisted monolayers.

We studied the manifestation of reconstruction effects through piezoelectric charge networks, which generate electrical potentials on the material surface that strongly depend on the strain configuration. In addition to dislocations, whose potential profiles reveal the strain orientation relative to the zigzag and armchair directions, piezopotential peaks were observed at the network nodes. These peaks act as charge defects, with their majority charge sign determined by the strain configuration and twist angle. By defining a form factor for the potential peaks, we identify their characteristic wave vectors as a function of the deformation in the upper layer.

Our results indicate that, while hBN exhibits excellent properties as a substrate for van der Waals heterostructures, strain during fabrication can induce inhomogeneous electric potentials. Consequently, when using hBN as a substrate or tunnel barrier, the potential perturbation of charge carriers must be considered, rather than treating it as a passive element in the structure.

\section*{Acknowledgements}

I.S. acknowledges financial support from the University of Manchester's Dean's Doctoral Scholarship. This work was supported by EPSRC Grants EP/S030719/1 and EP/V007033/1, and the Lloyd Register Foundation Nanotechnology Grant. J.G.M. is supported by the University of Manchester Dame Kathleen Ollerenshaw Fellowship. The authors acknowledge the use of resources provided by the Isambard 3 Tier-2 HPC Facility, hosted by the University of Bristol and operated by the GW4 Alliance (\url{https://gw4.ac.uk}) and funded by UK Research and Innovation; and the Engineering and Physical Sciences Research Council [EP/X039137/1].

\bibliography{references}

\appendix

\section{Lattice reconstruction of strained
moir{\'e} superlattices}\label{Append:LatticeRelaxation}

\begin{figure}
    \centering
    \includegraphics[width=0.8\linewidth]{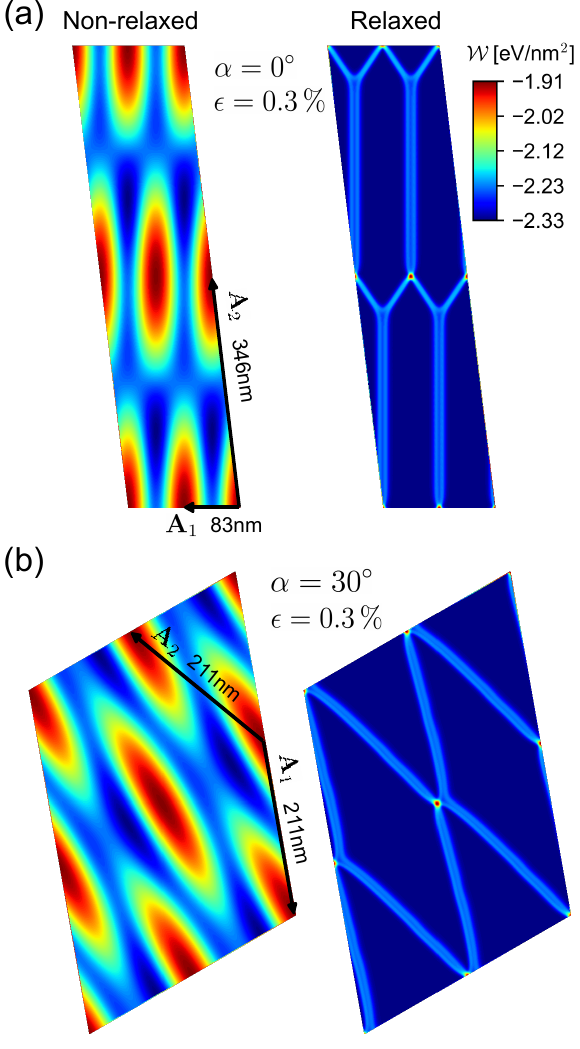}
    \caption{Adhesion energy maps for non-relaxed (left panel) and relaxed (right panel) moir{\'e} superlattices formed from the stacking of a $\epsilon=0.3\%$ uniaxially strained hBN monolayer on top of an eight-layer film with $\theta=0^{\circ}$. Strain is applied along the (a) zigzag ($\alpha=0^{\circ}$) and (b) armchair direction ($\alpha=30^{\circ}$).}
    \label{fig:RelaxedSuperlattice}
\end{figure}

The optimal configuration of displacement fields for strained and/or twisted monolayers stacked on a multilayer hBN film is found considering the energy of the moir{\'e} supercell (mSC) defined by vectors $\bvec{A}_{1}$ and $\bvec{A}_{2}$ (Eq. \ref{eq:moirevectors}),
\begin{equation}
\begin{split}
    \mathcal{E}_{\rm mSC} = \int_{\rm mSC} \D ^{2}r\, \Bigg[ &\mathcal{W}_{1',1}(\rr_0^{(1',1)}) + \mathcal{B}_{1} + \mathcal{B}_{1'}\\
    &+ \sum_{\ell'=2'}^{M'} \Big\{ \mathcal{W}_{\ell'-1,\ell'}(\rr_0^{(\ell'-1,\ell')}) + \mathcal{B}_{\ell'} \Big\} \Bigg],
\end{split}
\end{equation}
where the adhesion and elastic energy densities are defined in Sec. \ref{Sec:Model}. The lateral offset vector at the twisted/strained interface is determined by \eqref{Eq:StackingStrain}, whereas in the $M$ layer film as Eq. \eqref{Eq:StackingBulk}. Fig. \ref{fig:RelaxedSuperlattice} shows the resultant adhesion energy density at the interface for two representative scenarios of a $\epsilon=0.3\%$ strained monolayer on top of an eight-layer film ($M=8$) with twist angle $\theta=0^{\circ}$.

\section{Strain magnitude and twist angle dependence of the piezoelectric potential centres}\label{Append:StrainTwistPotential}

\begin{figure}[h!]
    \centering
    \includegraphics[width=0.95\linewidth]{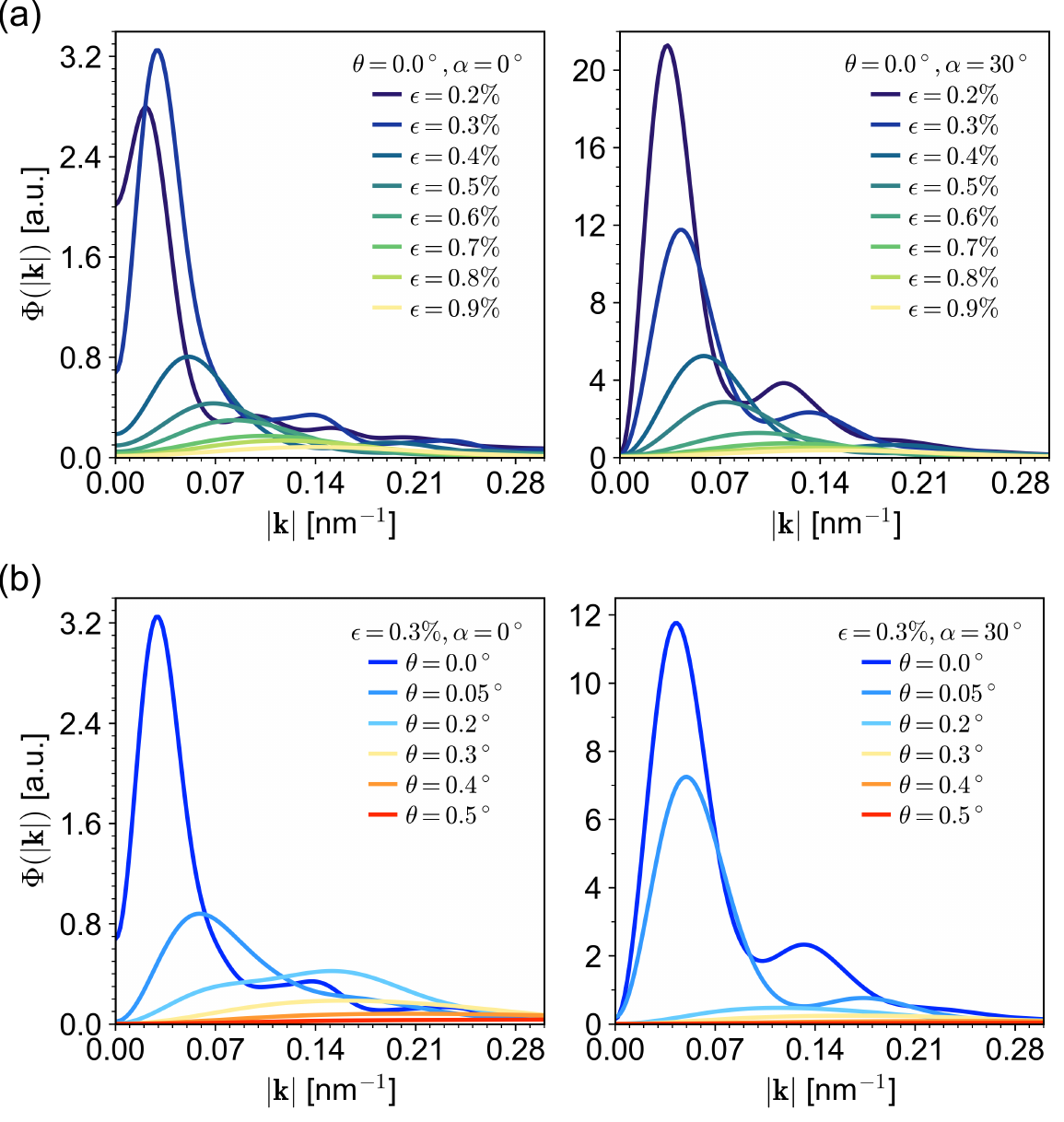}
    \caption{Strain magnitude and twist angle dependence of the piezopotential peak form factor for a uniaxially strained monolayer along zigzag (left panels) and armchair (right panels) stacked on bulk hBN. (a) Wave-vector dependence for different magnitudes of strain in a non-twisted top monolayer. (b) Form factor in a $\epsilon=0.3\%$ strained top layer for different twist angles.}
    \label{fig:strainTwist}
\end{figure}

We present the strain magnitude dependence of the potential peak form factor $\Phi(|\kk|)$ at the dislocation network nodes. For this, we consider the consider non-twisted top monolayers strained along the zigzag ($\alpha=0^{\circ}$) and armchair ($\alpha=30^{\circ}$) direction. Fig. \ref{fig:strainTwist}(a) shows the form factor for different values of strain $\epsilon<1\%$. In both cases, the main peaks observed shows a gradual decrease and broadening as strain increases, reflecting on the decline of the reconstruction effects.

We also analyse the twist angle evolution of the potential peaks for a given configuration of strain in the top layer. Considering a $\epsilon=0.3\%$ strained monolayer, the presence of sharp peaks is restricted to the regime $\theta[{\rm rad}]<\epsilon$ ($\theta\lesssim 0.17^{\circ}$ in this case), as shown in Fig. \ref{fig:strainTwist}(b).

\end{document}